\documentclass{eas}
\usepackage{graphicx}
\begin{document}


\title{Phase masks in astronomy: \\From the Mach-Zehnder interferometer 
to Coronographs } \runningtitle{Dohlen: Phase masks in astronomy 
\dots}
\author{Kjetil Dohlen}
\address{Laboratoire d'Astrophysique de Marseille \\
2 place Leverrier, 13248 Marseille Cedex 4, FRANCE \\ Tel : + 33 
(0) 4 95 04 41 24, Fax : + 33 (0) 4 91 62 11 90 \\ e-mail : 
kjetil.dohlen@oamp.fr}
\begin{abstract}
Phase masks have numerous applications in astronomical optics, in 
particular related to two themes: coronography for detection and 
analysis of extrasolar planets or circumstellar disks, and 
wavefront analysis for extremely precise adaptive optics systems 
or cophasing of segmented mirrors. I review some of the 
literature concerning phase masks and attempt to bridge the gap 
between two instrumental systems in which they are often found: 
the Mach-Zehnder interferometer and the coronograph.
\end{abstract}
\maketitle
\section{Introduction}
Zernike (\cite{Zernike}) introduced phase masks into astronomical 
optics when he  proposed the use of a phase-shifting spot to 
replace Foucault's classical knife edge for testing of optical 
surfaces. He showed that this allowed photometric measurement of 
surface phase errors rather than surface slope errors, and, as a 
bi-product of this work, he found it convenient to introduce a 
certain series of polynomials, facilitating the analysis of his 
surface maps. Zernike's name is forever associated with the 
polynomials rather than with the phase mask. A similar procedure 
was later proposed by Smartt \& Strong (\cite{Smartt72}). 

\begin{figure}[htb]
  \centering
  \includegraphics[width=6cm]{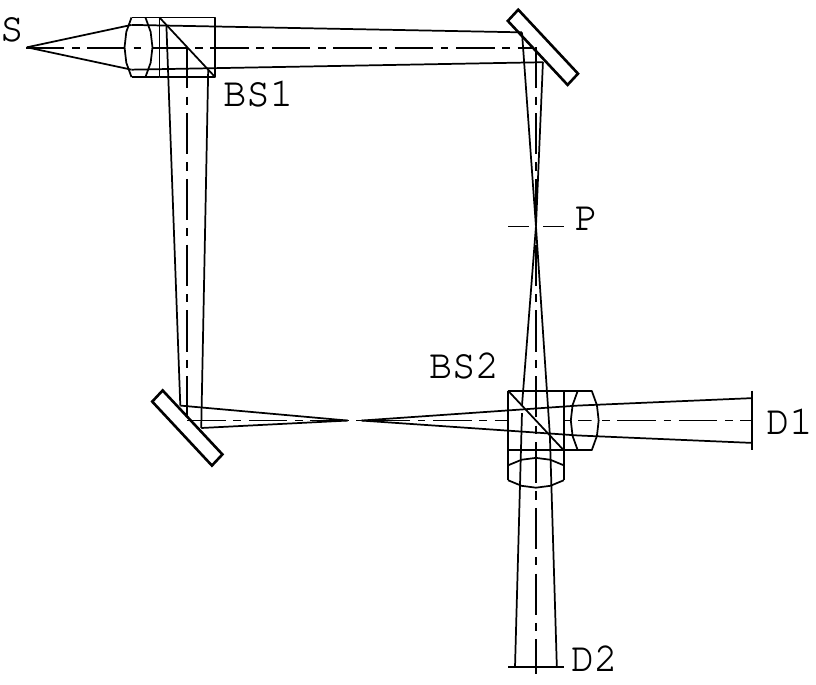}
  \caption{\label{MZschema}Diagram of the Mach-Zehnder interferometer 
  as a wavefront sensor with a pinhole (P) in one of the arms in an 
  image of the source (S).}
\end{figure}

Angel (\cite{Ang94}) proposed to use an adaptation of the Zernike 
test based on a Mach-Zehnder interferometer as wavefront sensor 
for high-performance adaptive optics systems. The Mach-Zehnder 
interferometer has the advantage over the Michelson 
interferometer of separating the output beam from the input beam, 
hence providing one-way traffic between separation and 
recombination of the beams, see Fig. \ref{MZschema}. Letting the 
beam go through a focus within the interferometer, beam 
modification by spatial filtering is possible, and Angel's 
proposition was to filter the beam in one arm by a pin-hole the 
size of an Airy disk, thus creating a spherical reference wave. 

More recently, the same basic principle has been proposed as a 
phasing sensor for extremely large telescopes (Dohlen \etal\ 
\cite{Kona98}). The pinhole is here larger, roughly the size of 
the atmospheric seeing disk, in order to minimize the effect of 
atmospheric turbulence on the estimation of inter-segment phase 
errors in segmented mirrors.

Another astronomical optical component resembling Zernike's test 
is the Roddier phase mask coronograph (Roddier \& Roddier 
\cite{Roddier97}), providing interferometric extinction of 
starlight in the hope of achieving direct imagery of extrasolar 
planets. But while Zernike's dot was adjusted to produce a 
roughly 90$^{\circ}$ phase shift, Roddier's dot needs to produce 
a very precise 180$^{\circ}$ phase shift. Also, where the former 
was fairly tolerant to dot size, the latter required a diameter 
equal to a precise fraction of the Airy disk. Of course, as the 
dot's phase shifting capacity is chromatically dependent unless 
composed of a stack of dielectric materials, and since the Airy 
disk's diameter also varies with wavelength, the required nulling 
can only be achieved within very narrow wavelength bands.

We have recently proposed a generalization of Roddier's 
coronograph where the dot is surrounded by a ring with a 
different phase shift (Soummer \etal\ \cite{Soummer03}). As 
wavelength increases, the phase lag decreases both for the dot 
and the ring, but since at the same time the fractions of the 
Airy disk covered also vary, complex vector analysis shows that 
the total transmission remains close to zero over a wide 
wavelength range, making this component an interesting option for 
wide-band stellar coronography.

In this paper we describe fundamentals and practical aspects of 
these instruments, illustrated by simulations. In Section 2, we 
describe the major functionalities of the Mach-Zehnder 
interferometer in its astronomical sensor applications, and in 
Section 3 we present a vector-based discussion of (circular) 
phase-mask coronographs. In Section 4 we investigate the 
equivalence between the Mach-Zehnder interferometer and phase 
mask coronographs and indicate situations in which the use of one 
or the other may be beneficial from a performance and/or 
practical implementation point of view.

\section{The Mach-Zehnder interferometer}
The Mach-Zehnder interferometer serves as a base concept for 
numerous instruments within all fields of optics. Astronomical 
optics is no exception as illustrated for example by the imaging 
Fourier transform spectrometer currently under construction for 
the SPIRE instrument (Swinyard \etal\ \cite{Spire02}) for ESA's 
3.5-meter diameter far infrared space telescope Herschel 
(Pilbratt \cite{Pil01}). We are here concerned with rather 
different astronomical applications, where the goal is not to 
analyze the spectral content of the source, but to analyze the 
shape of the transmitted wavefront in order to correct it 
adaptively or actively.

\subsection{Wavefront sensor for extreme AO}

In the concept proposed by Angel (\cite{Ang94}), the wavefront is 
split into two equal parts just before going through a focus, see 
Figure 1. In one of the arms, a pinhole placed in the focus acts 
as a spatial filter and provides a spherical reference wavefront. 
At the output beamsplitter, the original wavefront interferes 
with the reference wavefront, and fringes representing the 
wavefront defects are observed in a pupil image. Since the 
Mach-Zehnder interferometer has two accessible outputs, 
containing complementary fringe patters, two images are formed; 
the difference between these images is referred to as the 
Mach-Zehnder signal.

\begin{figure}[htb]
  \centering
  \includegraphics[width=8cm]{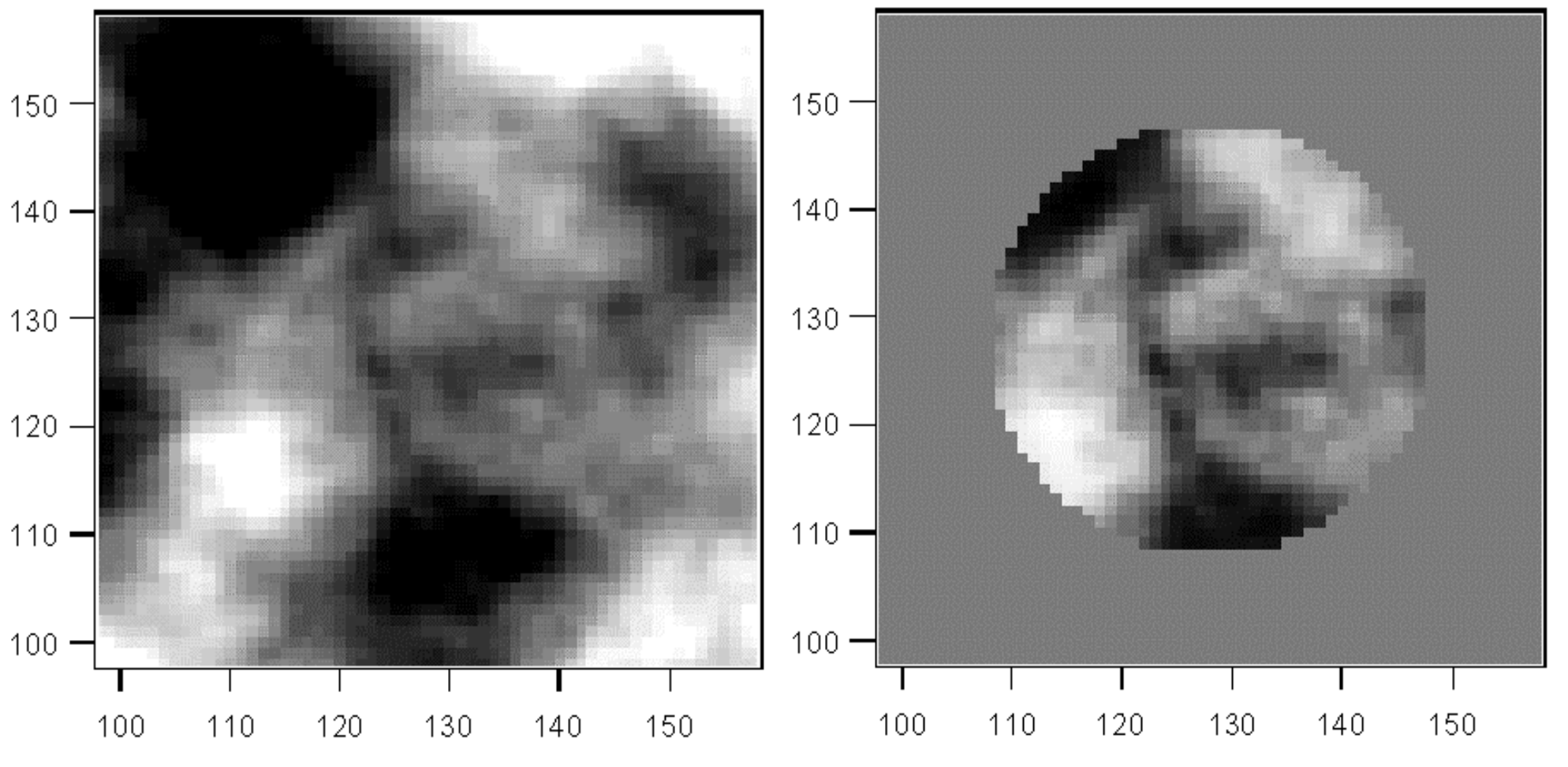}
  \caption{\label{MZsim}Phase screen representing a wavefront error 
  of $0.1\lambda$ rms (left) and the measured Mach-Zehnder signal 
  for an OPD between the two arms of $\lambda/4$ (right).}
\end{figure}

\begin{figure}[htb]
  \centering
  \includegraphics[width=12cm]{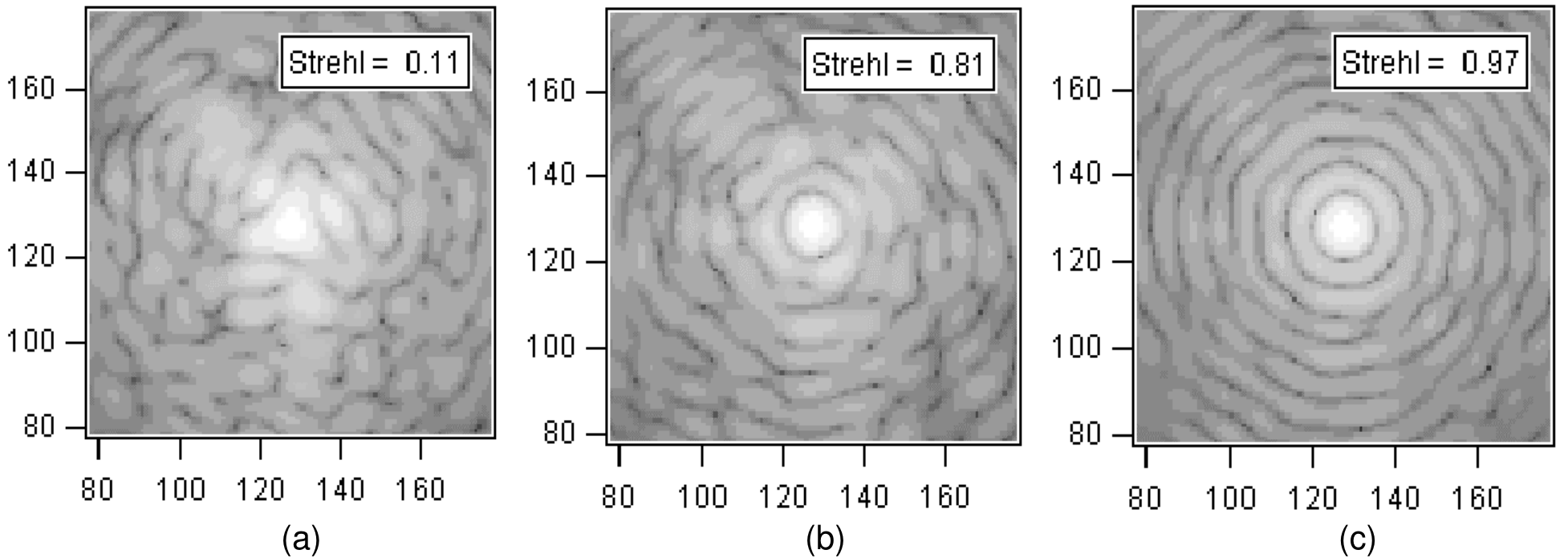}
  \caption{\label{MZloop}PSF corresponding to the first three steps of 
  simulated closed loop operation of a Mach-Zehnder-based 
  adaptive optics system,
  assuming linear reconstruction and ignoring all noise sources. 
  Intensity is coded logarithmically.}
\end{figure}

When the optical path difference (OPD) between the arms is 
adjusted to zero, the signal is proportional to the cosine of the 
wavefront errors, ie, for small errors, the signal is related to 
the square of the wavefront error. If, however, the OPD is set to 
$\lambda/4$, then the signal is proportional to the sine of the 
wavefront error, hence linear for small errors. 
Reconstruction-free wavefront sensing is therefore possible in 
closed loop operation, if a one-to-one relationship exists 
between the pixels of the signal and the deformable mirror 
actuators. This is illustrated in Fig. \ref{MZsim} where the 
phase screen representing an atmospheric wavefront of 
$0.1\lambda$ rms error is compared with the measured signal. 
Assuming a linear approximation for the sine, the wavefront 
estimation error is in this case $0.025\lambda$, but in closed 
loop, this residual is rapidly reduced, as illustrated in Fig. 
\ref{MZloop}, showing the PSF for  the first three steps of 
simulated closed loop operation assuming linear reconstruction 
and ignoring all noise sources.

Stahl \& Sandler \cite{Stahl} investigated the use of this type 
of wavefront sensor in their study of an adaptive system for 
exoplanet detection consisting of an 11~000-actuator DM 
controlled at a 2 kHz update rate. They concluded that, while 
instrumentally challenging, Jupiter-like planet could be detected 
around nearby stars, and that a survey of 30 stars within 10 pc 
could be reasonably foreseen using the 6.5m MMT.

Laboratory demonstrations of very fine-scale wavefront retrieval 
has been achieved (Langlois \etal\ \cite{Langlois}), where the 
Mach-Zehnder signal was captured using two 128x128 pixel CMOS 
detector arrays and pixel-by-pixel coupled with a liquid crystal 
(LC) spatial phase modulator. While alignment between detectors 
and LC limited the performance, a residual phase error smaller 
than 1/5 of the $\lambda/18$ initial wavefront error was reported.

This method works fine for small wavefront errors, but it is less 
adapted to errors exceeding $\lambda/4$ ($\pi/2$ radians), as is 
generally the case for the uncorrected atmosphere. Three options 
allow to overcome this problem: Pre-correction using classical 
low-order AO, rapid phase shifting using a piezo actuator to vary 
the OPD between the Mach-Zehnder arms, and boot-strapping by 
actively varying the pinhole size. The first option is 
conceptually simple and allows for upgrading of existing AO 
systems, but the total system becomes rather complex. The second 
option, assumed by Stahl \& Sandler (\cite{Stahl}) requires 
extremely high measurement rates to allow sequential phase 
stepping in the presence of an evolving phase screen. Also, the 
flux transmitted by the pinhole is very small compared to the 
total wavefront flux, giving a low-contrast interferogram. 

\begin{figure}[htb]
  \centering
  \includegraphics[width=12cm]{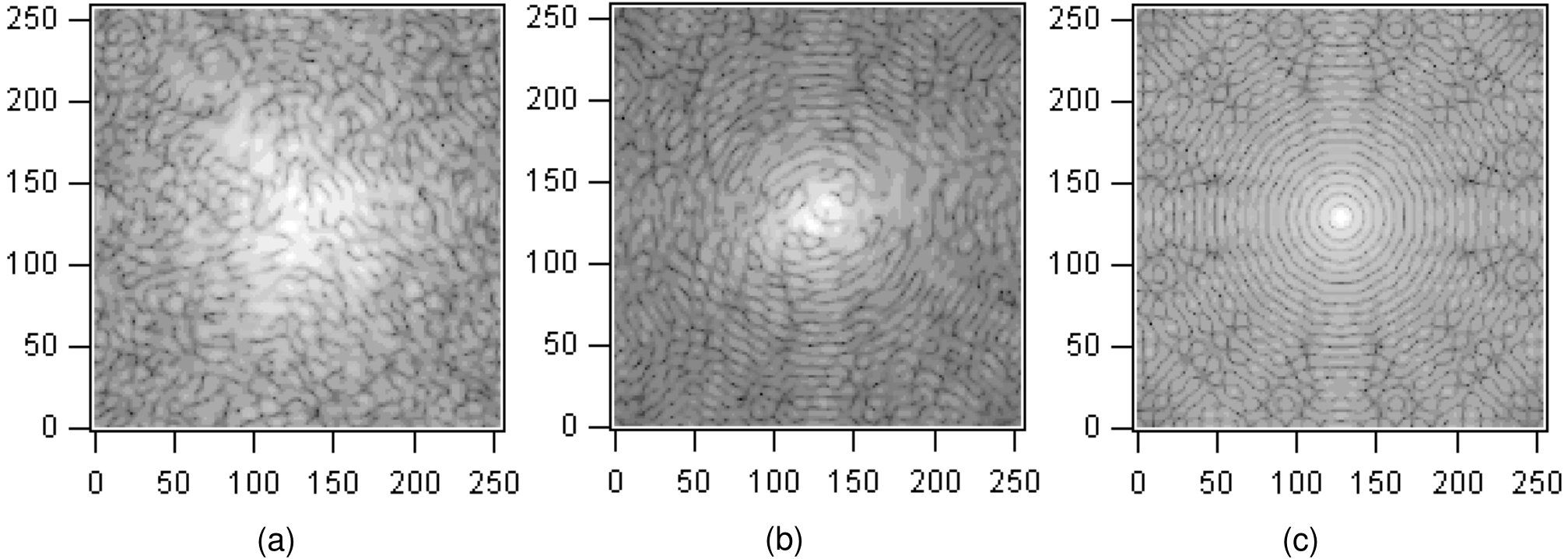}
  \caption{\label{MZboot}PSFs representing the starting point (a), an 
  intermediate step (b), and the final result (c) for a boot-strapped 
  Mach-Zehnder wavefront sensor whose pin-hole diameter varies 
  dynamically. Intensity is coded logarithmically.}
\end{figure}

The boot-strapping method, mentioned by Angel (\cite{Ang94}), is 
certainly the most elegant option, although the fabrication of a 
variable-size precision pinhole may be a challenge. Increasing 
the pinhole, which acts as a spatial low-pass filter, shifts the 
cut-off frequency upwards. After recombination, the resulting 
wavefront represents the difference between the filtered and 
unfiltered wavefronts, and so the total operation can be seen as 
high-pass filtering, transmitting only the wavefront components 
beyond the cut-off frequency given by the pin-hole size. For a 
pinhole size corresponding to the seeing disk, the remaining 
aberrations can be expected to represent approximately 1 radian, 
appropriate for the reconstruction-free approach. Once closed 
loop operation has been established, the pinhole size can 
gradually be reduced until the diffraction limit is reached. 
Figure \ref{MZboot} shows the starting point PSF, an 
intermediate, and the final corrected PSF for a very simple 
simulation using a fixed phase screen whose initial wavefront rms 
was $\lambda/2$ and a dynamically variable pinhole diameter. 

\subsection{Cophasing sensor}
The use of a Mach-Zehnder with a large pinhole has also been 
proposed in the context of measuring inter-segment phase steps in 
extremely large telescopes (ELTs) (Dohlen \etal \ \cite{Kona98}). 
All current ELT projects depend on the use of segmented primary 
mirrors, and some projects (Dierickx \& Gilmozzi \cite{Owl2000}) 
also rely on a (flat) segmented secondary mirror. While the basic 
technologies required for segmented telescopes have been 
demonstrated on the Keck telescopes, ELTs of diameters from 30 to 
100 m represent quantum leaps of one to two orders of magnitude 
in segmentation complexity compared with current 10-m technology. 

Calibration of the phasing system remains an important issue, and 
it is not clear that the proven Keck technology, based on 
accurate positioning of micro pupils at the intersections between 
segments, is transposable to the ELT case. A general issue 
concerns the accuracy required in micro pupil mask fabrication 
and alignment, a more specific issue, related to the use of two 
segmented mirrors, is that of separating the two segmentation 
patters. The latter point is aggravated by the fact that guide 
stars used for phasing measurements can be located anywhere 
within the technical field of view (FOV) of the telescope. As the 
guide star moves across the FOV, the segmentation patterns move 
one with respect to the other, making the use of a fixed micro 
pupil mask impossible.

An alternative approach consists of coding the phase errors as 
intensity variations in a pupil image projected onto a detector 
array. This allows direct pixel-by-pixel processing and avoids 
the use of a pupil mask. Several methods have been proposed to 
achieve intensity coding of the phase errors, including the use 
of pupil defocus, first proposed and demonstrated by Chanan 
\etal\ \cite{Chanan99} for the Keck telescope and later described 
theoretically in terms of wave front curvature by Cuevas \etal\ 
(\cite{Cuevas}) in analogy with curvature sensing for adaptive 
optics (Roddier \cite{Roddier90}). More recently, it has been 
shown that the pyramid wave front sensor---a modernized version 
of the Foucault knife-edge test, developed for adaptive optics---
is also sensitive to segment dephasing (Esposito \etal\ 
\cite{Pyramid}). Use of the Mach-Zehnder interferometer was first 
proposed by Dohlen \etal\ (\cite{Kona98}), and described in its 
present form by Montoya Martinez \etal\ (\cite{Hawaii02}).   

\begin{figure}[htb]
  \centering
  \includegraphics[width=12cm]{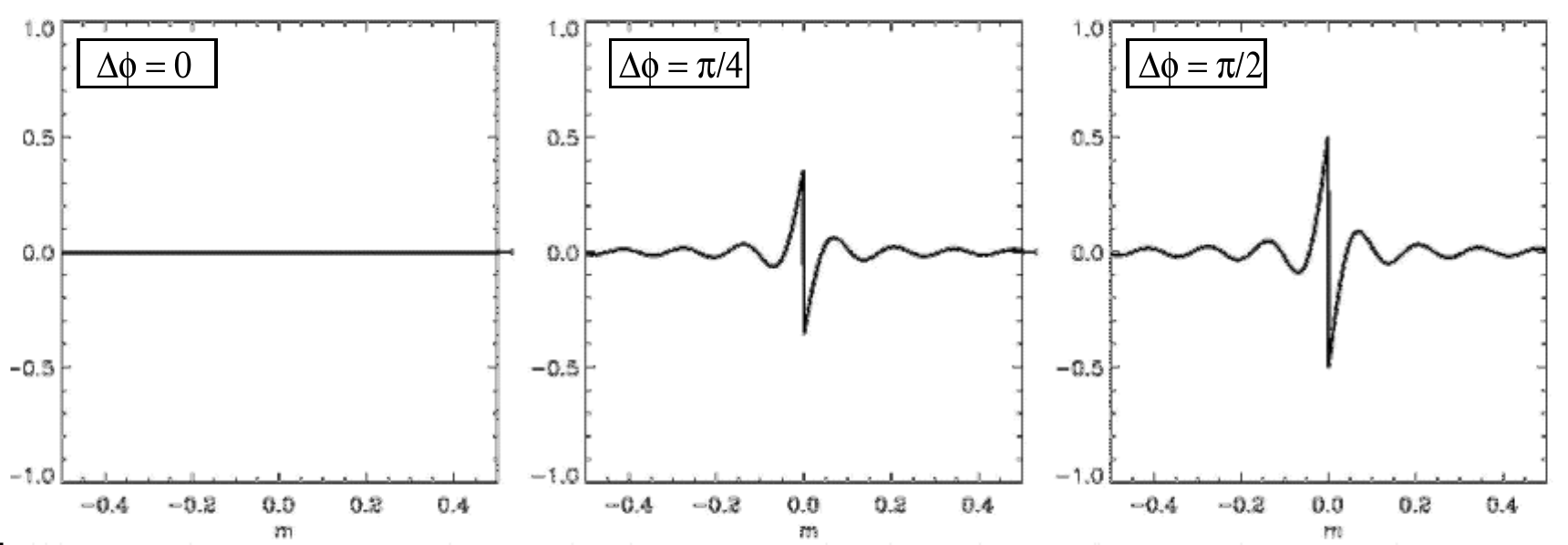}
  \caption{\label{MZsignal}Mach-Zehnder signal in the absence of atmosphere 
  for piston values of 0 (left), $\pi/4$ (middle), and $\pi/2$ 
  (right).}
\end{figure}

As in the bootstrapping procedure described above, the low 
frequencies of the atmospheric errors are removed by using a 
pinhole approximately the size of the seeing disk. This also 
removes the low-frequency components of the segment phase errors, 
but since the frequency content of a step function is very wide, 
sufficient information remains. As seen in Fig. \ref{MZsignal}, 
showing the signal profile for different piston values in the 
absence of atmospheric turbulence, a brightening on one side and 
a darkening on the other side of the edge appears, with a sharp 
transition between dark and bright. The dark-bright amplitude is 
proportional to the sine of the piston step, with a peak at 
$\pi/2$ and falling to zero at $\pi$. The signal is perfectly 
symmetrical, and identical for a pistons of $\pi/2\pm\delta$.

\begin{figure}[htb]
  \centering
  \includegraphics[width=12cm]{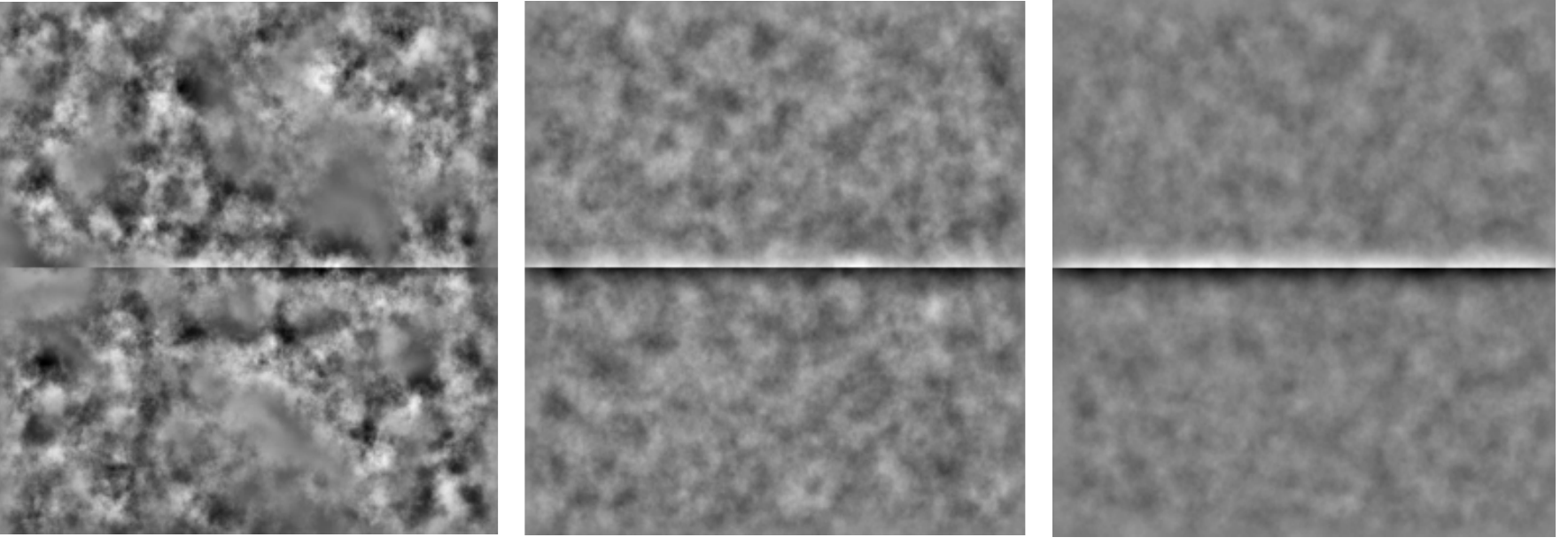}
  \caption{\label{EdgeAtm}Simulated Mach-Zehnder signal for different 
  exposure times, representing one (left), 25 (middle) and 100 (right) 
  phase screens.}
\end{figure}

Since the high spatial frequency atmospheric phase residuals vary 
with time, they tend to disappear in long-exposure, see Fig. 
\ref{EdgeAtm}. Optimal performance requires exposure times of 
typically a few seconds.

\begin{figure}[htb]
  \centering
  \includegraphics[width=12cm]{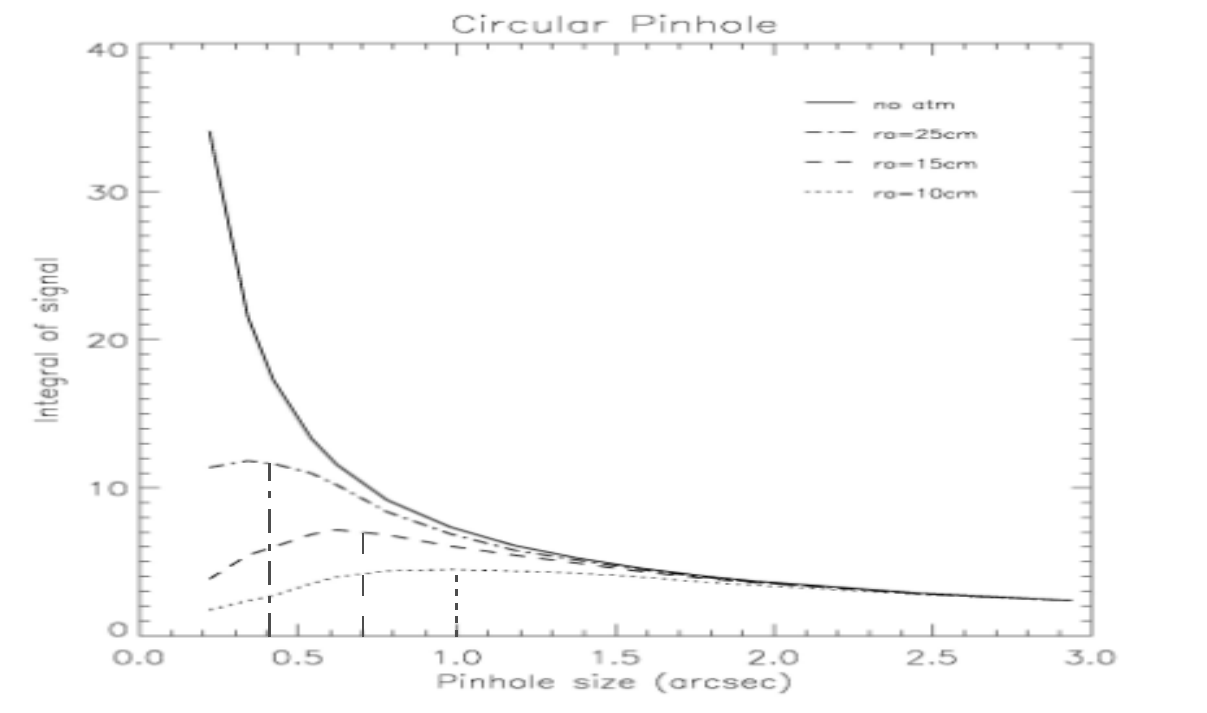}
  \caption{\label{MZcal}Calibration curves for the Mach-Zehnder 
  phasing sensor in the absence of atmosphere (solid line) and for 
  different atmospheric conditions. Vertical bars indicate the 
  seeing disk dimension.}
\end{figure}

Extracting the piston information from the Mach-Zehnder signal 
can be done by estimating the dark-to-bright amplitude, or as the 
difference between the integral of the signal between the edge 
and the first zero crossing on either side of the edge. Fig. 
\ref{MZcal} shows the result of the integral method for a piston 
of $\pi/2$ as a function of pinhole size in the absence of 
atmospheric errors (continuous line) and for three different 
atmospheres. While the signal amplitude stays constant whatever 
the pinhole, the width of the signal, and hence the integral, 
diminishes as the pinhole grows. In the presence of atmospheric 
errors, the signal disappears for small pinholes, goes through a 
maximum close to the seeing disc diameter (indicated by vertical 
bars), and joins the ideal curve for pinholes larger than about 
twice the seeing disc diameter. 

The presence of gaps and turned-down edges, as well as practical 
problems related to signal detection involving pixel convolution 
and sampling, affects the signal significantly. In particular, 
effects of turned-down edges such as the zero-piston residual 
signal, are amplified as the pinhole size grows, indicating the 
use of small pinholes. Optimal performance is in practice 
obtained by using a pinhole approximately the size of the 
median-seeing seeing disk associated with an iterative alignment 
procedure. While the use of a small pinhole results in an 
uncertainty in the exact piston amplitude because the seeing 
parameter is variable and {\em a priori} unknown, it minimizes 
the zero-offset problem induced by turned-down edges. Simulations 
indicate that an error below 20nm after a few iterations will be 
achieved.

\section{Phase mask coronographs}
\begin{figure}[htb]
  \centering
  \includegraphics[width=12cm]{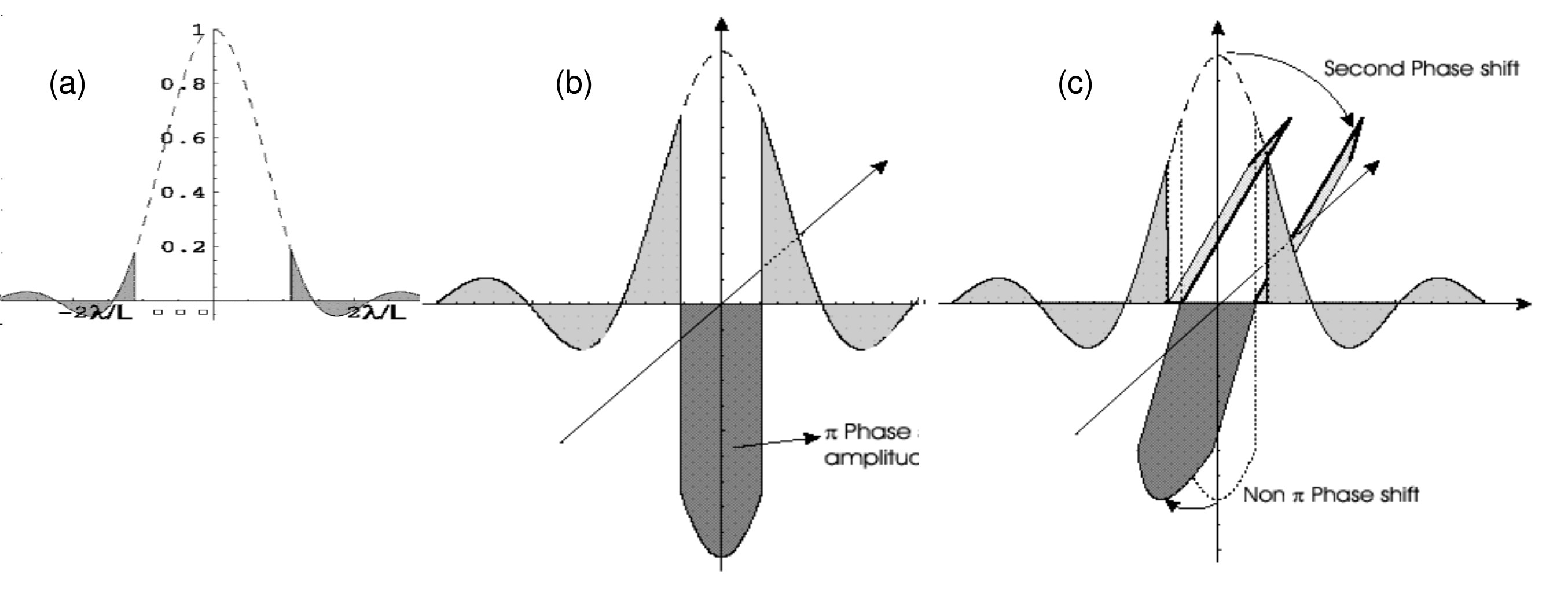}
  \caption{\label{PhaseCoro}Illustration of the principle of 
  interferential nulling phase mask coronographs: Lyot (a), 
  Roddier (b) and dual zone (c). In each case, the integral 
  of the complex field transmitted through the mask is zero.}
\end{figure}

While early stellar coronography for imaging of fait objects 
close to bright stars were simple adaptations of the Lyot solar 
coronograph, the phase mask coronograph proposed by Roddier \& 
Roddier (\cite{Roddier97}) was the first to utilize the spatial 
coherence of stellar sources and to obtain nulling by 
interferometric extinction. Instead of blocking the stellar flux, 
part of it was phase shifted by 180$^{\circ}$ in order to 
interfere destructively with the remaining flux. A simple 
illustration of this technique based on Fourier theory can be 
made by considering the center of the pupil, where the value of 
the complex electric field is equal to the integral of the 
image-plane electric field. In order to bring the field at the 
pupil center to zero, the size of the phase mask must be adjusted 
so that the integral within the phase mask exactly matches the 
integral outside the mask, see Fig. \ref{PhaseCoro} (b). 
Elsewhere in the pupil, the nulling is not perfect, and a careful 
optimization of the mask diameter will be necessary to select the 
optimum nulling zone. However, perfect nulling can be achieved by 
this method if the entrance pupil is apodized (Guyon \& Roddier 
\cite{Guyon2000}, Aime \etal\ \cite{Aime2002}).

The Lyot stellar coronograph can also be used in a nulling mode 
by adjusting the diameter of the (opaque) mask such that the 
integral of the transmitted flux is zero, see Fig. 
\ref{PhaseCoro} (a). Again, apodization is required for optimal 
nulling (Aime \etal\ \cite{Aime2002}). This concept is referred 
to as the Prolate Apodized Lyot Coronograph (PALC).

\begin{figure}[htb]
  \centering
  \includegraphics[width=12cm]{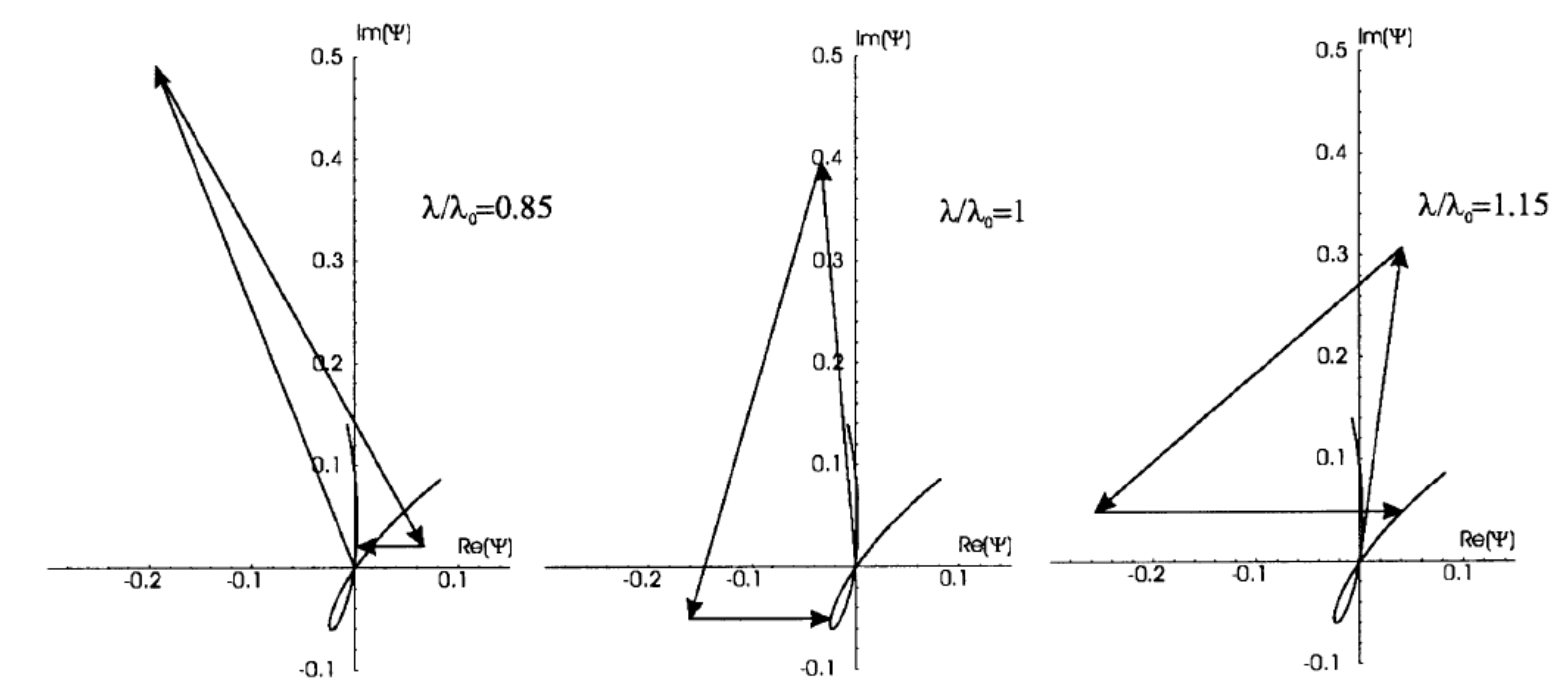}
  \caption{\label{DZvectors}Illustration of the three-vector 
  addition in the complex plane. The resulting vector describes a 
  $\gamma$-like curve as the wavelength varies across the band.}
\end{figure}

Both the Roddier and Lyot nulling coronograph are chromatic, with 
optimal performance at a single wavelength. At other wavelengths, 
the size of the diffraction spot changes, and so the value of the 
integrals. Also, for the Roddier concept, the optical path 
through the phase mask changes unless built up as an achromatic 
combination of several materials. Soummer \etal\ 
(\cite{Soummer03}) has proposed an achromatized, dual zone (DZ) 
version of the phase-mask coronograph, where the phase shifting 
mask is surrounded by a phase shifting annulus, see Fig. 
\ref{PhaseCoro} (c). An infinite number of solutions can be found 
for which the complex integral of the field transmitted by this 
mask is zero, providing zero amplitude at the center of the exit 
pupil at a given wavelength. By fairly simple analysis, one can 
also show that solutions exist which give zero integral at two 
different wavelengths, as illustrated in Fig. \ref{DZvectors} in 
terms of complex vector summation. Optimal nulling over the 
entire pupil and for a given band is found by numerical least 
squares optimization of the diameter of each mask zone and their 
optical thicknesses as well as apodization parameters. It is 
interesting to note that this concept is improved by applying 
complex apodization, involving the variation of phase as well as 
transmission across the pupil. In practice, phase apodization is 
achieved simply by introducing a slight defocus of the 
coronograph mask.

\section{Mach-Zehnder--coronograph equivalences}
The Mach-Zehnder interferometer combined with a pinhole and a 
phase shifter as described above is similar to a phase mask 
coronograph. In fact, each Mach-Zehnder output has an equivalent 
coronographic phase mask of complex transmission ${\bf M}_{\rm 
coro} = {\bf M}_{1} \pm {\bf M}_{2}$, where ${\bf M}_{1}$ and $ 
{\bf M}_{2}$ are complex transmissions of the focal plane mask in 
each Mach-Zehnder arm and the choice of sign corresponds to the 
choice of output. In the MZ systems we have considered so far for 
wavefront sensing, ${\bf M}_{1}$ has unit transmission everywhere 
(transparent) and  ${\bf M}_{2}$ has unit transmission within the 
pinhole and zero outside. The optical path difference introduced 
between the arms is accounted for by adding a phase angle to one 
of the mask functions. 

\begin{figure}[htb]
  \centering
  \includegraphics[width=12cm]{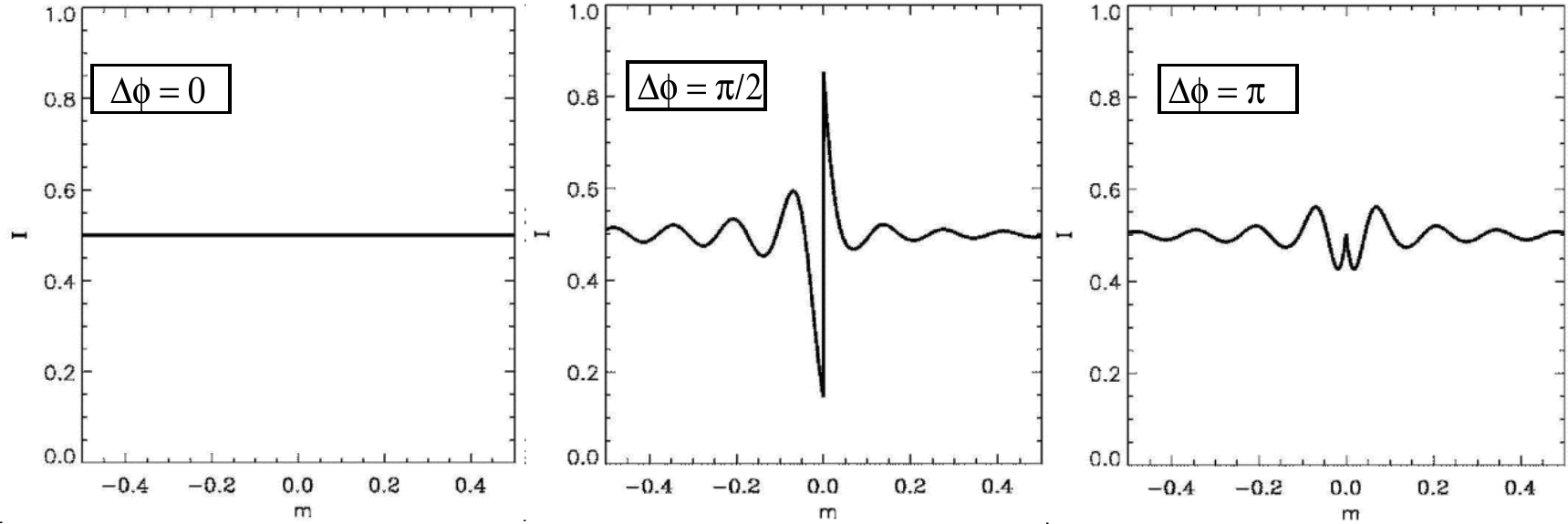}
  \caption{\label{MZcorosignal}Signal profiles for the coronographic 
  version of the Mach-Zehnder phasing sensor in the absence of atmosphere 
  and for piston values of 0 (left), $\pi/2$ (middle), and $\pi$ 
  (right). The mask has unit transmission and a 45$^{circ}$ dephasing.}
\end{figure}

A possibility offered by this equivalence is to replace the 
interferometer in the phasing sensor by its coronographic 
equivalence. Although the advantage of having access to two 
outputs in terms of signal cleanliness etc. is lost, this clearly 
would simplify the opto-mechanical design of the system. Strictly 
speaking, the coronographic equivalence of the proposed 
Mach-Zehnder phasing sensor has a 45$^{\circ}$ phase-shifting 
spot with twice the transmission of its surroundings. Studying 
different phase and transmission distributions, it has been found 
that en equal-transmission mask with a 45$^{\circ}$ phase 
shifting dot gives an optimally symmetrical signal. Fig. 
\ref{MZcorosignal} shows the signal profile obtained in this case 
for piston steps of 0, $\pi/2$ and $\pi$. The two leftmost 
profiles can be compared with the left and right profiles of Fig. 
\ref{MZsignal}, respectively. A slight asymmetry develops as the 
piston step grows, and at piston of $\pi$, where the 
``classical'' Mach-Zehnder signal is zero, a small residual is 
present. This can actually be seen as an advantage, allowing an 
increase of the calibration range. 

On the other hand, interferometric equivalences of coronographic 
devices may offer access to a larger design space than the simple 
phase mask approach. Also, the concept can be generalized by 
using transmission phase masks rather than pinholes, and 
different masks in each arm. Playing with mask material 
(dispersion, absorption, etc) and dimensions, a large number of 
variables are available, allowing optimization of performance 
according to applications and requirements. In particular, this 
could provide coronographs of the dual zone type while avoiding 
the difficulty of manufacturing the dual zone mask.

Other interesting hybrids involving coronographs and Mach-Zehnder 
interferometers are being published at the time of this writing 
by Codona \& Angel (\cite{CodAng}) and Labeyrie (\cite{Lab04}), 
both of which are based on active interferometric speckle 
suppression (``active halo nulling''). The star image is focussed 
onto the first beam splitter (BS1 in Fig. \ref{MZschema}), which 
is replaced by a mirror with a pinhole approximately the size of 
the Airy disk. This acts as a Lyot-type coronograph for the 
reflected beam, which, in conjunction with appropriate 
apodization, provides an image ridded of the Airy pattern but 
polluted by a residual speckle halo. The transmitted beam, 
containing the major part of the spatially coherent starlight, is 
modified, either using phase and amplitude spatial light 
modulators (Codona \& Angel \cite{CodAng}) or a real-time  
generated holographic optical element (Labeyrie \cite{Lab04}), 
and made to interfere, destructively, with the coronographic beam 
at the output beam splitter. The value of this approach is 
reported to be the relaxed precision required for the wavefront 
control of the reference beam, greatly reduced compared with the 
requirements for the pre-coronograph wavefront control.

\section{Conclusion}
Phase masks are used in various branches of astronomical optics, 
in particular for wavefront characterization and coronographic 
nulling. We have reviewed different concepts and applications of 
phase masks both in their coronographic form and in the form of 
Mach-Zehnder interferometers. While phase mask coronographs will 
be important in the search and characterization of extra-solar 
planets and circumstellar disks, the Mach-Zehnder wavefront 
sensor will certainly have an important role to play in future 
high-performance adaptive optics systems and as a phasing sensor 
for future generations of large segmented telescopes. 

The equivalence between coronographic and interferometric phase 
masks has been established, and possible advantages of using one 
or the other in terms of performance and feasibility have been 
pointed out. Finally, extremely interesting developments of 
Mach-Zehnder--coronograph hybrids using  active interferometric 
speckle suppression have been reported.

 \vspace{0.75cm}

{\bf Acknowledgements} I am grateful to Luzma Montoya Martinez 
for providing illustrations from her work on the Mach-Zehnder 
phasing sensor, and to Antoine Labeyrie for inviting me to make 
this talk as a {\em College de France} seminar.


\end{document}